\documentclass[a4paper,aps,prl,preprintnumbers,amsmath,amssymb,twocolumn]{revtex4}
\usepackage{bbm}
\usepackage{amsmath}
\usepackage{verbatim}
\usepackage{graphicx}
\usepackage{epsfig}

\newcommand{\id}{{\rm id}}
\newcommand{\tr}[1]{{\rm tr}\left[{#1}\right]}

\newcommand{\be}{\begin{equation}}
\newcommand{\ee}{\end{equation}}
\newcommand{\bea}{\begin{eqnarray}}
\newcommand{\eea}{\end{eqnarray}}

\begin{document}
%\title{Gaussian states have extremal entanglement}
\title{Quantum Capacities of Bosonic Channels}

\author{Michael M. Wolf$^{1}$, David P\'erez-Garc\'ia$^{2}$, Geza Giedke$^{1}$} \affiliation{$^{1}$ Max-Planck-Institute for Quantum
Optics, Hans-Kopfermann-Str.\ 1, D-85748 Garching,
Germany.\\$^{2}$ \'Area de Matem\'atica Aplicada, Universidad Rey
Juan Carlos, C/ Tulipan s/n, 28933 M\'ostoles (Madrid), Spain}

\begin{abstract}
We investigate the capacity of bosonic quantum channels for the
transmission of quantum information. Achievable rates are
determined from measurable moments of the channel by showing that
every channel can asymptotically simulate a Gaussian channel which
is characterized by second moments of the initial channel. We
calculate the quantum capacity for a class of Gaussian channels,
including channels describing optical fibers with photon losses,
by proving that Gaussian encodings are optimal. Along the way we
provide a complete characterization of degradable Gaussian
channels and those arising from teleportation protocols.
\end{abstract}

\date{\today}

\maketitle

One of the aims of \emph{Quantum Information Theory} \cite{NC} is
to follow the ideas of Shannon and to establish a theory of
information based on the rules of quantum mechanics. A key problem
along this way is the calculation of the quantum capacity of noisy
quantum channels. That is, the question how much quantum
information---measured in number of \emph{qubits}---can be
transmitted reliably per use of a given channel? Despite
substantial progress \cite{Qcap} this can be answered only in very
few  cases \cite{DS} as a simple formula, comparable to Shannon's
coding theorem, is not known.

In this work we investigate the quantum capacity of \emph{bosonic
channels} which might
 describe transmission in space, as light sent through optical fibers, or in
time, like in quantum memories \cite{Qmemory}. The paper has two
parts. In the first part we prove that the quantum capacity of any
bosonic channel $T$ is lower bounded by that of a corresponding
Gaussian channel $T_G$, which can be derived from measurable
moments of $T$. This implies that for determining and certifying
achievable rates for the transmission of quantum information
through $T$ we need not know the channel exactly (which might be
hardly possible in infinite dimensions), but merely its second
moments, i.e., a few measurable parameters. In the second part we
then explicitly calculate the quantum capacity of a class of
Gaussian channels, which includes the important case of
attenuation channels modelling optical fibers with photon losses
and broad-band channels where losses and photon number constraints
might be frequency dependent. Along the way we provide two tools
that might be of independent interest: a complete characterization
of degradable Gaussian channels and of those arising from
teleporting through Gaussian states.

\section{Preliminaries}

Before we derive the main results  we will briefly recall the
basic notions \cite{HolevoBook,JWGauss}. Consider a bosonic system
of $N$ modes characterized by $N$ pairs of canonical operators
$(Q_1,P_1,\ldots,Q_N,P_N)=:R$ for which the commutation relations
$[R_k,R_l]=i\sigma_{kl}$ are governed by the \emph{symplectic
matrix} $\sigma$. The exponentials $W_\xi:=e^{i\xi
R},\;\xi\in\mathbb{R}^{2N}$ are called \emph{Weyl displacement
operators}. Their expectation value, the \emph{characteristic
function}, $\chi(\xi):=\tr{\rho W_\xi}$ is the Fourier transform
of the Wigner function and for \emph{Gaussian states} it is a
Gaussian \be \chi(\xi)=e^{i \xi\cdot d -
\frac14\xi\cdot\Gamma\xi}\;,\ee with first moments $d_k=\tr{\rho
R_k}$ and \emph{covariance matrix} (CM)
$\Gamma_{kl}:=\tr{\rho\{R_k-d_k,R_l-d_l\}_+}$. Note that coherent,
squeezed and thermal states in quantum optics are all Gaussian
states.

\emph{Gaussian channels} \cite{JWGauss,Lindblad} transform Weyl
operators as $W_\xi\mapsto W_{X\xi}e^{-\frac14 \xi Y\xi}$
 and act on
covariance matrices as \be\gamma\mapsto X^T\gamma
X+Y\;.\label{XY}\ee Particularly important instances of
single-mode Gaussian channels are attenuation and amplification
channels for which $X=\sqrt{\eta}\mathbbm{1}$ and
$Y=|\eta-1|\mathbbm{1}$. For $0\leq\eta\leq 1$ this models a
single mode of an optical fiber with transmissivity $\eta$ where
the environment is assumed to be in the vacuum state. The latter
reflects the fact that thermal photons with optical frequencies
are negligible at room temperature. For $\eta> 1$ the channel
becomes an amplification channel, where the noise term $Y$ is now
a consequence of the Heisenberg uncertainty.\vspace{3pt}

{\bf Teleportation channels:} We will now derive the form of
Gaussian channels which are obtained when teleporting through a
centered bipartite Gaussian state. As this is useful for applying
but not necessary for understanding the following it might be
skipped by the reader.
Let $\Gamma={\footnotesize \left(%
\begin{array}{cc}
  \Gamma_A & \Gamma_C \\
  \Gamma_C^T & \Gamma_B \\
\end{array}%
\right)}$ be the CM of a Gaussian state of $N_A+N_B$ modes with
$N_A=N_B$. Assume Bob wants to teleport a quantum state of $N_B$
modes with CM $\gamma$ to Alice. Using the standard protocol
\cite{teleV} he sends pairs of modes from $\gamma$ and $\Gamma_B$
through 50:50 beam-splitters, measures the $Q$ and $P$
quadratures, and then communicates the outcomes. Depending on the
latter Alice applies displacements to the modes in $\Gamma_A$. The
simplest way of deriving an expression for the output is to start
with the Wigner representation and to assume that the state to be
teleported is a centered Gaussian. The Wigner function before the
measurement is up to normalization given by $\exp{-\xi\big[
M_{BS}^T(\Gamma\oplus\gamma)^{-1} M_{BS}\big]\xi}$, where $M_{BS}$
corresponds to the beam-splitter operation. With
$\xi=(\xi_A,\xi_B,\xi_{B'})$ the final Wigner function is then
proportional to \be\label{Wigner}\int d\xi_B d\xi_{B'}
e^{-\xi\big[ M_X^TM_{BS}^T(\Gamma\oplus\gamma)^{-1}
M_{BS}M_X\big]\xi}\;,\ee where $M_X$ incorporates the
displacements, i.e., it is the identity matrix plus an arbitrary
$2 N_B\times 2 N_B$ off-diagonal block which maps the $2 N_B$
measurement outcomes onto the respective displacements. In order
to circumvent integrating Eq.(\ref{Wigner}) we can now go to the
characteristic function, i.e., the Fourier transformed picture.
The integration then boils down to picking out the upper left
block of the inverted matrix $\big[
M_X^TM_{BS}^T(\Gamma\oplus\gamma)^{-1} M_{BS}M_X\big]^{-1}$. The
inversion is, however, trivial since $M_{BS}^{-1}=M_{BS}^T$ and
$M_X^{-1}$ is obtained from $M_X$ by changing the sign of all
off-diagonal entries. In this way we obtain that the input CM is
transformed to \bea\label{telechannel}\gamma\ \mapsto\ X^T\gamma
X\ + && \Big[\Gamma_A+ \Gamma_C\Lambda X + (\Gamma_C\Lambda X)^T\\
&& \ \ + X^T\Lambda^T\Gamma_B\Lambda X\Big]\nonumber\;,\eea where
$\Lambda=\text{diag}(1,-1,1,-1,\ldots)$ and $X$ is such that
$\sqrt{2} X$ is the matrix of displacement transformations, i.e.,
the \emph{gain} which is typically chosen to be
$\sqrt{2}\mathbbm{1}$.

Clearly, Eq.(\ref{telechannel}) has the form (\ref{XY}) and
following the above lines it is straight forward to show that the
channel is  Gaussian and maps any (not necessarily centered
Gaussian) input characteristic function $\chi_{in}$ into
\be\chi_{out}(\xi)=\chi_{in}(X \xi)\chi_\Gamma(\xi\oplus\Lambda
X\xi)\label{telechi}\;.\ee For standard protocols
($X=\mathbbm{1}$) on single modes ($N_A=N_B=1$) this was derived
in \cite{Scutaru}.

%Note that if $\Gamma$ is a collection
%of two-mode squeezed states, i.e., $\Gamma_A=\Gamma_B= \mathbbm{1}
%\cosh r, \Gamma_C= -\Lambda \sinh r$, then for
%$r\rightarrow\infty$ and $X=\mathbbm{1}$ this becomes the ideal
%channel.

\section{Achievable rates for arbitrary channels}

The subject of interest is the quantum capacity $Q(T)$ of an
arbitrary---a priori unknown---channel $T$. We will show how one
can certify achievable rates for the transmission of quantum
information through $T$ by only looking at the CM $\Gamma$ of a
state $\rho_T=(T\otimes\id)(\psi)$ which is obtained by sending
half of an arbitrary entangled state $\psi$ through the channel.
 $\Gamma$ could be determined by homodyne measurements. The argument combines (i) the relation between entanglement
distillation and quantum capacities observed in \cite{Bennett},
(ii) the extremality of Gaussian states shown in \cite{CLT} and
(iii) the explicit form of Gaussian teleportation channels derived
in the previous section. All together this leads to the chain of
inequalities \be Q(T)\geq D_\leftarrow(\rho_T)\geq
D_\leftarrow({\cal G}(\rho_T))\geq Q(T_G)\label{main1}\;.\ee Here
$D_\leftarrow(\rho_T)$ is the distillable entanglement under
protocols with one-way communication (from Bob to Alice). Since a
classical side channel does not increase $Q(T)$ this is clearly a
lower bound to the capacity as Alice and Bob could simply first
distill $\rho_T$ and then use the obtained maximally entangled
states for teleportation \cite{Bennett}. The second inequality
uses that replacing $\rho_T$ by a Gaussian state ${\cal
G}(\rho_T)$ with the same CM $\Gamma$ can only decrease the
distillable entanglement \cite{CLT}. Finally, if we use the
Gaussian state in turn as a resource for establishing a
teleportation channel $T_G$ we end up with the sought inequality
$Q(T)\geq Q(T_G)$. $T_G$ is then the Gaussian channel in
Eqs.(\ref{telechannel},\ref{telechi}), which is for a fixed
teleportation protocol (a fixed matrix $X$) completely  determined
by $\Gamma$.

\begin{figure}[t]
\begin{center}
\epsfig{file=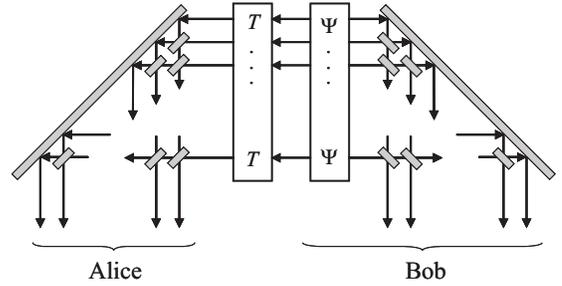,angle=0,width=0.84\linewidth}
\end{center}
\caption {In order to obtain a Gaussian channel from an arbitrary
quantum channel $T$ Bob (the sender) prepares $n$ instances of an
entangled state $\psi$ half of which he sends through $T^{\otimes
n}$. After applying two arrays of 50:50 beam-splitters to the
output $\rho_T^{\otimes n}=[(T \otimes \id)(\psi)]^{\otimes n}$
the $n$ reduced states will converge to a Gaussian state ${\cal
G}(\rho_T)$ (with the same CM as $\rho_T$) which can in turn be
used to establish a Gaussian teleportation channel $T_G$.}
\label{figarray}
\end{figure}

 Bounds on the quantum capacity
of Gaussian channels were derived in \cite{HolWer,GioShor} and we
will show below that it can be calculated exactly for some
important cases. Note that a simple bound for $Q(T)$ can be
obtained from a lower bound to $D_\leftarrow({\cal G}(\rho_{T}))$,
the conditional entropy of the Gaussian state with CM $\Gamma$,
i.e., $Q(T)\geq S(\Gamma_A)-S(\Gamma)$.

Before we proceed, two comments on the quality of the above bound
and its operational meaning are in order: The given argument holds
for arbitrary $T$ and $\psi$. However, since we bound by Gaussian
quantities the inequality might become trivial (i.e., $Q(T_G)= 0$
though $Q(T)\gg 0$) if both $T$ and $\psi$ are too far from being
Gaussian. On the other hand, if $T$ is Gaussian and
$|\psi\rangle=(\cosh r)^{-1}\sum_n (\tanh r)^n |nn\rangle$ is a
two-mode squeezed state, then in the limit $r\rightarrow\infty$
the inequality becomes tight, i.e., $Q(T_G)\rightarrow Q(T)$ with
 exponentially vanishing gap.

The first and last step in Eq.(\ref{main1}) have a simple
operational meaning: sender and receiver first establish some
entanglement, distill it and then use it as a resource for
teleportation. The second step can also be understood in
operational terms. To achieve $\rho_T\mapsto {\cal G}(\rho_T)$
both sender and receiver have to apply an array of beam-splitters
(see Fig.\ref{figarray}) to many copies of the shared state
$\rho_T$. Asymptotically every reduced state at the output will
then converge to ${\cal G}(\rho_T)$ \cite{CLT}. By applying this
\emph{gaussification} to many different subsets, sender and
receiver can then distill from or teleport through independent
copies of ${\cal G}(\rho_T)$ and in this way asymptotically
simulate the channel $T_G$ via $T$.

\section{Quantum capacity of Gaussian channels}

It was proven in \cite{Qcap} that the quantum capacity of a
quantum channel $T$ can be expressed as \bea Q(T) &=&
\lim_{n\rightarrow\infty}\frac1n \sup_{\rho} J\big(\rho,T^{\otimes
n}\big)\;,\\
J(\rho,T) &=& S\big(T(\rho)\big) - S\big((T\otimes
\id)(\psi)\big)\;,\eea where $\psi$ is a purification of $\rho$
and $J$ is known as the \emph{coherent information}. In general,
the calculation of $Q(T)$ from the above formula is a daunting
task since (i) the coherent information is known to be not
additive, i.e., the regularization $n\rightarrow\infty$ is
necessary, and (ii) due to lacking concavity properties there are
local maxima which are not global ones. On top of this, for
bosonic channels the optimization is over an an infinite
dimensional space.

Fortunately, for a class of Gaussian channels including the
important case of the lossy channel, these obstacles can be
circumvented by exploiting recent results on degradability of
channels \cite{DS,Gio2} and extremality of Gaussian states
\cite{CLT}.

To this end consider a channel $T(\rho_S)={\rm tr}_E[U
(\rho_S\otimes\varphi_E) U^\dag]$ expressed in terms of a unitary
coupling between the system $S$ and the environment $E$ which is
initially in a pure state $\varphi_E$. The \emph{conjugate
channel} $T_c(\rho_S)={\rm tr}_S[U (\rho_S\otimes\varphi_E)
U^\dag]$ is defined as a mapping from the system to the
environment. As shown in \cite{DS} the coherent information can be
expressed in terms of a conditional entropy if there exists a
channel $T'$ such that $T'\circ T=T_c$ --- in this case $T$ is
called \emph{degradable}. More precisely, if $\tilde{\rho}_{S'E'}$
is the extension of the state $\tilde{\rho}_{S'}=T'\circ T (\rho)$
to the environment $E'$ of $T'$, then \be\label{condent}
J(\rho,T)=S(\tilde{\rho}_{S'E'})-S(\tilde{\rho}_{S'})=:S(E'|S')\;.\ee
The conditional entropy $S(E'|S')$ is known to be strongly
sub-additive \cite{NC}, i.e., for a composite system
$S(E_{12}'|S_{12}')\leq S(E_{1}'|S_{1}')+S(E_{2}'|S_{2}')$. This
has important consequences: for a set $\{T_i\}$ of degradable
channels $J(\rho,\otimes_i T_i)\leq \sum_i J(\rho_i,T_i)$, where
$\rho_i$ are the corresponding reduced states, and if each $T_i$
is a Gaussian channel,  we have in addition \be J(\rho,\otimes_i
T_i)\leq\sum_i J(\rho_i,T_i)\leq \sum_i J({\cal
G}(\rho_i),T_i).\ee The last inequality follows from the
extremality of Gaussian states w.r.t. the conditional entropy
\cite{JWGauss,CLT} together with the fact that for Gaussian
channels $T_c$ can be chosen to be Gaussian and the CM is
transformed irrespective of whether the input was Gaussian or not.
\begin{figure}[t]
\begin{center}
\epsfig{file=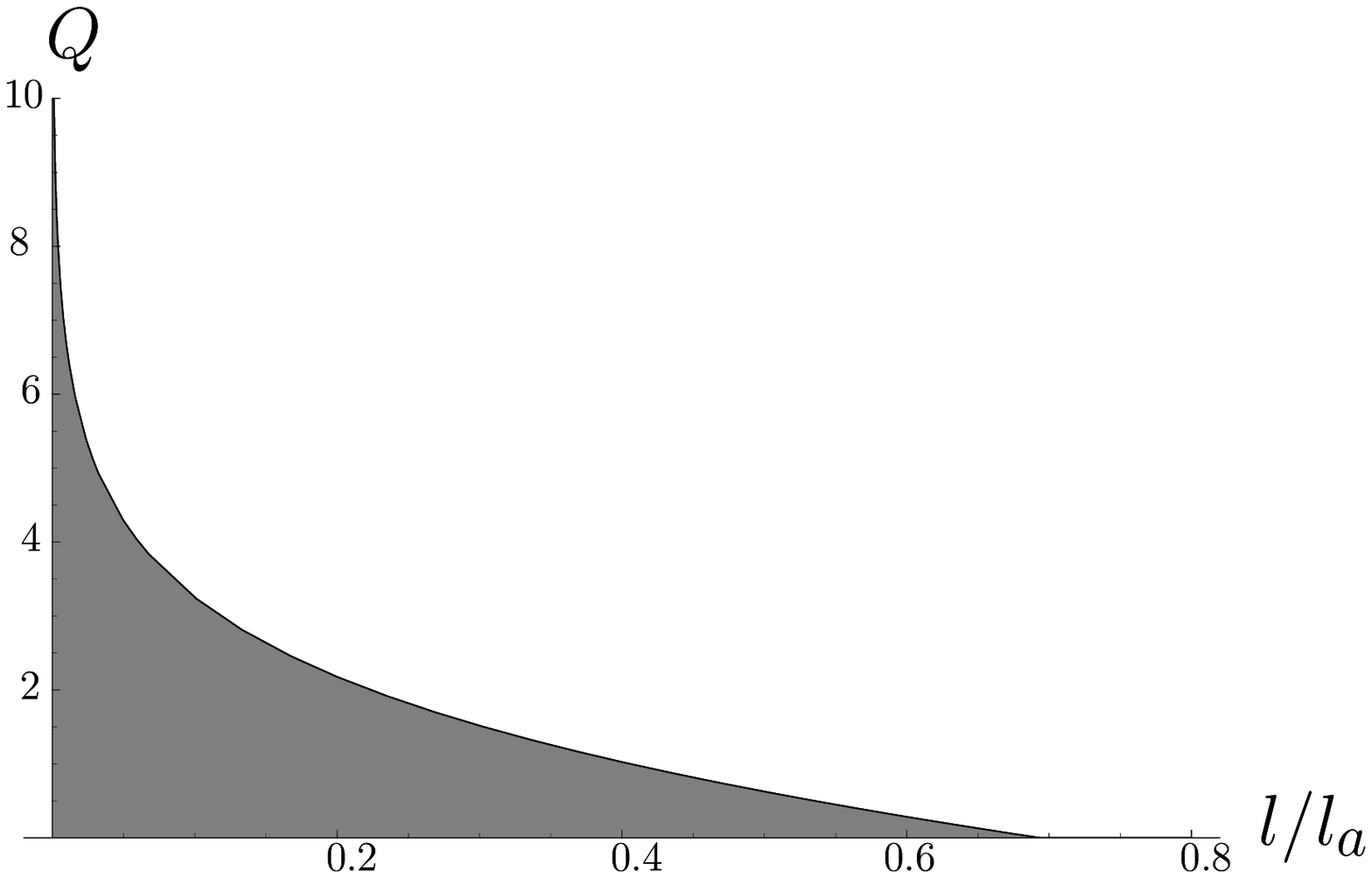,angle=0,width=0.84\linewidth}
\end{center}
\caption {\label{fig2} Quantum capacity of a channel with photon
losses as a function of the transmission length $l$ in terms of
the absorbtion length $l_a$, i.e., $\eta=e^{-l/l_a}$. For quantum
memories $l$ and $l_a$ are storage and decay time. The capacity
vanishes for $l/l_a=\ln 2\approx 0.693$, where the channel can be
considered to be  part of a symmetric approximate cloning channel.
}
\end{figure}
As a consequence, if   $T_i$ are degradable Gaussian channels,
then \be Q\big(\otimes_i T_i\big)=\sum_i
\sup_{\rho_G}J\big(\rho_G,T_i\big)\;,\label{QG}\ee where the
supremum is now taken only over Gaussian input states $\rho_G$.
Calculating the latter for Gaussian channels is now a feasible
task which was solved for the single-mode case in \cite{HolWer}
and in \cite{GioShor} for broadband channels under power
constraints using Lagrange multipliers. In fact, if we impose a
constraint on the input energy of the form $\sum_i \omega_i
N_i={\cal E}$, where $N_i$ is the average input photon number of
mode $i$ with corresponding frequency $\omega_i$, then the above
argumentation still holds, since the constraint just depends on
the CM. The importance of Eq.(\ref{QG}) stems from the fact that a
large class of Gaussian channels is indeed degradable, as shown in
\cite{Gio2} and extended below. In particular we can apply
Eq.(\ref{QG}) to attenuation (amplification) channels with
transmissivity $\eta$ (gain $\sqrt{\eta}$). Together with the
optimization carried out in \cite{HolWer} this yields $Q$ as a
function of $\eta$ (see Fig.2): \be\label{Qeta}
Q(\eta)=\max\big\{0,\log_2 |\eta| -\log_2 |1-\eta|\big\}\;. \ee
%This
%vanishes for $\eta\leq\frac12$, which can be seen as a direct
%consequence of the no-cloning theorem.
Note that the quantum capacity of every degradable Gaussian
channel can easily be calculated as $J$ becomes a concave function
of the CM such that local maxima are global ones. This is again a
direct consequence of Eq.(\ref{condent}) together with the
concavity of the conditional entropy \cite{NC} and the extremality
of Gaussian states \cite{JWGauss,CLT}.

\section{Degradable Gaussian channels}

 We will now investigate the condition under which Eq.(\ref{QG})
 was derived and characterize the set of degradable Gaussian
 channels---extending the results of \cite{Gio2}. To this end we represent the channel in terms of a
 unitary coupling  between the system with $N_S$ modes and a (minimally represented)
 environment of $N_E\leq 2N_S$ modes which are initially in the vacuum state with $CM$
 $\gamma_E=\mathbbm{1}$. The interaction is described by a
 symplectic matrix of size $2(N_E+N_S)\times 2(N_E+N_S)$ which we write in block form as $S={\footnotesize\left(%
\begin{array}{cc}
  A & B \\
  C & D \\
\end{array}%
\right)}$ \cite{sympcond}. The output CM of the channel
$T:\gamma\mapsto X\gamma X^T+Y$ is then simply the lower right
block of $S(\gamma_E\oplus\gamma)S^T$ (i.e., $D=X$ and $C\gamma_E
C^T=Y$) whereas the conjugate channel $T_c$ corresponds to the
upper left block.

Let us first focus on the case $N_S=N_E$ and assume for simplicity
that the blocks in $S$ are non-singular. A channel is degradable
if $T_c\circ T^{-1}$ is completely positive which is for a
Gaussian trace preserving map equivalent to the condition
\cite{Lindblad} \be Y+i\sigma \geq i X\sigma X^T\;.\ee Inserting
the above block structure and using \cite{sympcond} shows that
complete positivity of $T_c\circ T^{-1}$ is equivalent to
\bea\label{skewEq}  0
&\leq& (\mathbbm{1}+i\sigma)-K(\mathbbm{1}+i\sigma)K^T\;,\\
&& K = C^TD^{-T}\sigma D^{-1} C\;.\nonumber\eea Expressing this in
terms of $X$ and $Y$ finally gives \cite{smallsteps}
\be\label{XYcond} \big(2 X\sigma X^T\sigma^T-\mathbbm{1}\big)Y\geq
0\;.\ee Similarly we can derive a condition for degradability of
$T_c$ (\emph{anti-degradability} of $T$) which is again given by
the expressions (\ref{skewEq}, \ref{XYcond}) which have then to be
negative instead of positive semi-definite.

Since for $N_E=N_S=1$ $X$ is a $2\times 2$ matrix and thus
$X\sigma X^T\sigma^T=\mathbbm{1}\det X$, condition (\ref{XYcond})
implies that either $T$ \emph{or} $T_c$ is degradable, as shown in
\cite{Gio2}. Hence, as anti-degradable channels have zero quantum
capacity (due to the no-cloning theorem), the quantum capacity of
every Gaussian channel with $N_S=N_E=1$ can easily be calculated.
In fact, by utilizing the freedom of acting unitarily before and
after the channel (which does not change its capacity) one can
bring the channel to a normal form \cite{SEW} which only depends
on the symplectic invariant $\det X$ such that $Q(T)$ of every
such channel is given by Eq.(\ref{Qeta}) with $\eta=\det X$.

Let us finally briefly comment on the case $N_E\neq N_S$.
%In
%principle one can follow the above lines and derive necessary and
%sufficient conditions for degradability in all cases. However,
%since this involves several technical case distinctions, we will
%merely sketch the main consequences.
If the environment is smaller than the system, then we can easily
follow the above lines for instance by choosing a representation
of the channel with larger $N_E$ equal to $N_S$
\cite{skewEqremark}. It is worth mentioning that if $S$
corresponds to a passive (i.e., number preserving) operation, then
for $N_E<N_S$ there are always unaffected modes such that
$Q(T)=\infty$ without additional constraints. If $N_E> N_S$ then
Eq.(\ref{XYcond}) is merely a necessary, whereas Eq.(\ref{skewEq})
is still a necessary and sufficient condition for degradability
\cite{skewEqremark}. Applying the latter to a general single-mode
channel with $N_S=1, N_E=2$ shows that generically one has neither
degradability nor anti-degradability. Hence, it remains open
whether in this case the capacity is given by Eq.(\ref{QG}).
However, we can easily derive an upper bound by exploiting the
fact that every Gaussian channel $T$ can be decomposed as
$T=T_1\circ T_2$, where $T_2$ is a  \emph{minimal noise channel}
\cite{Lindblad} for which $N_E=N_S$ with $X_2=X$, $Y_2\leq Y$ and
$T_1$ is a \emph{classical noise channel} for which
$X_1=\mathbbm{1}$, $Y_1=Y-Y_2$. Due to the bottleneck-inequality
for capacities (cf. \cite{HolWer}) we have $Q(T)\leq Q(T_2)$ where
the latter is in the single-mode case again given by
Eq.(\ref{Qeta}) with $\eta=\det X$. A lower bound is always given
by the r.h.s. of Eq.(\ref{QG}) as calculated in \cite{HolWer}.

\emph{In summary} we characterized the set of degradable Gaussian
channels and showed that their quantum capacity can be calculated
as it is attained for Gaussian product inputs. For arbitrary
non-Gaussian channels we derived a certifiable Gaussian lower
bound. Both ideas can be applied to finite dimensional systems as
well. This will, however, be the content of future work
\cite{future}.

\emph{Acknowledgements:} We appreciate valuable discussions with
J.I. Cirac and M.E. Shirokov.


\begin{thebibliography}{99}
\bibitem{smallsteps} To see this one can exploit the fact that $\frac12 (\mathbbm{1}\pm
i\sigma)$ are orthogonal projectors. In this way Eq.
(\ref{skewEq}) can be expressed in terms of the two conditions
$[\sigma,K]=0$ and $\mathbbm{1}\geq KK^T$ which can in turn be
rephrased
as \bea (X\sigma X^T\sigma Y) - (X\sigma X^T\sigma Y)^T &=& 0\;, \\
 (X\sigma X^T\sigma Y) + (X\sigma X^T\sigma Y)^T +Y &\leq& 0\;, \eea
 which together imply Eq.(\ref{XYcond}).
\bibitem{skewEqremark} The only assumption made in the derivation
of condition (\ref{skewEq}) is non-singularity of the diagonal
blocks $A$ and $D$.
\end{thebibliography}
\end{document}